\documentclass[twocolumn,amsmath,amssymb,prl,10pt,nofootinbib,superscriptaddress]{revtex4}
\usepackage{ graphicx, float,amsmath, amsmath, amssymb}
\usepackage[export]{adjustbox}

\def\be{\begin{equation}}
\def\ee{\end{equation}}
\def\bea{\begin{eqnarray}}
\def\eea{\end{eqnarray}}
\def\bse{\begin{subequations}}
\def\ese{\end{subequations}}

\usepackage[breaklinks, colorlinks, citecolor=blue]{hyperref}
\usepackage[labelsep=period]{caption}
\begin{document}
\title{Seeing through the cosmological bounce:\\
Footprints of the contracting phase and luminosity distance in bouncing models}
\author{Aur\'elien Barrau}%
\affiliation{%
Laboratoire de Physique Subatomique et de Cosmologie, Universit\'e Grenoble-Alpes, CNRS/IN2P3\\
53, avenue des Martyrs, 38026 Grenoble cedex, France
}
\author{Killian Martineau}%
\affiliation{%
Laboratoire de Physique Subatomique et de Cosmologie, Universit\'e Grenoble-Alpes, CNRS/IN2P3\\
53, avenue des Martyrs, 38026 Grenoble cedex, France
}
\author{Flora Moulin}%
\affiliation{%
Laboratoire de Physique Subatomique et de Cosmologie, Universit\'e Grenoble-Alpes, CNRS/IN2P3\\
53, avenue des Martyrs, 38026 Grenoble cedex, France
}
\date{\today}
\begin{abstract} 
The evolution of the luminosity distance in a contracting universe is studied. It is shown that for quite a lot of natural dynamical evolutions, its behavior is far from trivial and its value can even decrease with an increasing time interval between events. The consequences are investigated and it is underlined that this could both put stringent consistency conditions on bouncing models and open a new observational window on ``pre Big Bang" physics using standard gravitational waves.
 \end{abstract}
\maketitle

\section{Introduction}

The {\it Big Bang} is a prediction of general relativity (GR) in a regime where the theory is not valid anymore. Singularities are most probably pathologies of the models, not of spacetime itself. It is therefore natural to consider alternatives to the naive {\it Big Bang} image. Importantly, most models replacing the initial singularity by ``something else" were not designed to this aim but produce this desirable effect as a consequence of their application to the early universe (see \cite{Brandenberger:2016vhg,Battefeld:2014uga} and references therein for recent reviews). Among the countless ways to obtain a cosmological bounce, one can mention the null energy condition violation \cite{Peter:2001fy}, the strong energy condition violation with a positive curvature \cite{Falciano:2008gt}, ghost condensates \cite{Lin:2010pf}, galileons \cite{Qiu:2011cy}, S-branes\cite{Kounnas:2011fk}, quantum fields \cite{Cai:2007qw}, higher derivative terms \cite{Biswas:2005qr,Biswas:2006bs}, non-standard couplings \cite{Langlois:2013cya}, supergravity \cite{Koehn:2013upa}, and loop quantum cosmology \cite{Bojowald:2001xe,Ashtekar:2015dja}. These are only some examples among a much longer list which also includes, in a way, the ekpyrotic and cyclic scenarios \cite{Khoury:2001wf,Steinhardt:2001st}, together with string gaz cosmology \cite{Battefeld:2005av}. Bouncing models are natural extensions of the  {\it Big Bang} scenario and it comes as no surprise that they arise in many theories beyond GR. (Interestingly, those ideas are also being investigated in the black hole sector, see \cite{Barcelo:2017lnx} for a recent review). \\

All those models are obviously missing an observational confirmation or, at least, strong experimental constraints. As a legitimate step in this direction, many efforts were recently devoted to the calculation of primordial cosmological power spectra. Predictions for the cosmological microwave background (CMB) were made for nearly all the above-mentioned models (as exemples for specific settings, explaining the global strategy, one can consider \cite{Peter:2004um,Finelli:2007tr}).\\ 

In this article we follow another path. We investigate the unusual luminosity distance behavior in a contracting universe. We show that it is highly non-trivial. As a consequence, we raise some consistency issues for bouncing cosmological models. We finally suggest possible observational footprints of the contracting phase that could be observed through ``usual" gravitational waves.
 
\section{The luminosity distance in a contracting universe}

As far as observations are concerned, an important parameter is the luminosity distance $D_L$. It is defined by 
$
f=L/(4\pi D_L^2),
$
where $f$ is the observed flux from a given astrophysical source and $L$ is its luminosity. Intuitively, the luminosity distance is the ``equivalent" distance at which an object of the same luminosity should be in a usual euclidean space to lead to the same observed flux. In a flat expanding universe (in the presence of spatial curvature, the general expression involves trigonometric and hyperbolic fonctions \cite{Hogg:1999ad}), it reads as
\begin{equation}
D_L=c(1+z)\int_0^z\frac{dz'}{H(z')},
\label{DL2}
\end{equation}
where $H$ is the Hubble parameter and $z$ is the redshift. For our purpose, it is convenient to rewrite this formula as a function of time:
\begin{equation}
D_L=c(1+z)a(t_r)\int_{t_e}^{t_r}\frac{dt}{a(t)},
\label{DL3}
\end{equation}
where $t_e$ and $t_r$ are the emission and reception cosmic times of the considered signal and $a(t)$ is the sale factor. To study a contracting universe it is even better to get rid of the redshift and write the expression as
\begin{equation}
D_L=c\frac{a(t_r)^2}{a(t_e)}\int_{t_e}^{t_r}\frac{dt}{a(t)}.
\label{DL4}
\end{equation}

When one considers the contracting branch of a bouncing scenario, interesting and unusual phenomena can take place. Let us choose $t=0$ at the bounce time and assume that the universe contracts as $a(t)=k(-t)^n$ before the bounce (with $n=2/3$ for a matter-dominated phase and $n=1/2$ for a radiation-dominated phase). 
The  detailed evolution around the bounce could be {\it e.g.} given by the loop quantum cosmology modified Friedmann equation \cite{lqc9}
\begin{equation}
H^2=\frac{\kappa}{3}\rho\left(1-\frac{\rho}{\rho_c}\right),  \label{Friedman}
\end{equation}
where $\rho_c$ depends on the details of the model but can be guessed to be close to the Planck density.
However we have checked that the observables calculated in this article do not depend on the detailed shape of the modified equation of motion. We therefore approximate the scale factor by a constant function between $-t_B$ and $t_B$. Let $t_e$ and $t_r$ both be negative -- that is in the contracting branch -- with $t_e<t_r$. It is then easy to show that:
\begin{equation}
D_L=c\frac{(-t_r)^{2n}}{n-1}\left[ \frac{(-t_r)^{1-n}}{(-t_e)^n}-(-t_e)^{1-2n} \right].
\label{DL5}
\end{equation}
When $n<1/2$, $D_L\rightarrow \infty$ when $t_e\rightarrow -\infty$. This is in agreement with the intuitive behavior.\\

However, when $n>1/2$, $D_L\rightarrow 0$ when $t_e\rightarrow -\infty$. This is one of the important results we want to stress here. This strange behavior never happens in an expanding universe. It means that, for a fixed reception time $t_r$, an event that took place earlier in the contracting phase will be seen as {\it brighter}. Of course, the luminosity distance first increases with higher values of $-t_e$, reaches a maximum, and then decreases. The maximum can be shown (when $n \neq 1$) to be reached when
\begin{equation}
-t_e=\left[ \frac{2n-1}{n} \right]^{\frac{1}{n-1}}(-t_r).
\label{DLmax}
\end{equation}

When $n=1/2$, $D_L\rightarrow 2ca(t_r)^2/k^2$ when $t_e\rightarrow -\infty$. This means that events arbitrarily far away in the past will be detected at the same brightness once the asymptotic regime is reached.\\

\begin{figure}[H]
\includegraphics[width=85mm,center]{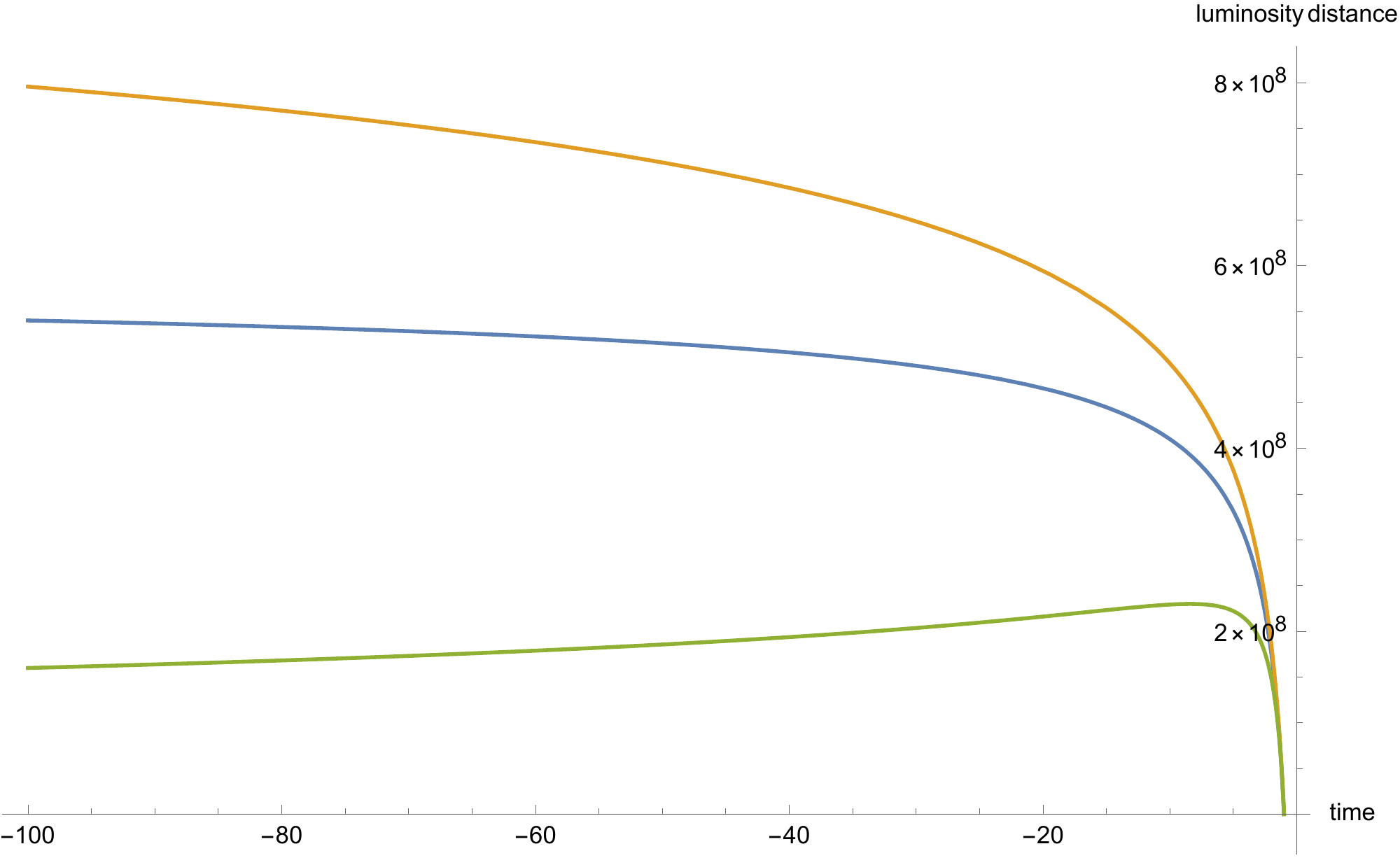}
\caption{Luminosity distance (m) as a function of the emission time $t_e$ (s) in the contracting branch for the power law contraction. The reception time $t_r$ has been set to 1 second before the bounce. The lower curve corresponds to $n=2/3$, the mid curve to $n=1/2$ and the upper curve to $n=0.45$.}
\label{alldl}
\end{figure}

Fig \ref{alldl} shows the luminosity distance evolution for three different values of $n$. It can be seen that $D_L$ is asymptotically constant in the remote past when $n=1/2$ and tends to 0 when $n>2/3$. The numerical values are not relevant and the plot aims at showing the global behavior. 

\begin{figure}[H]
\includegraphics[width=85mm,center]{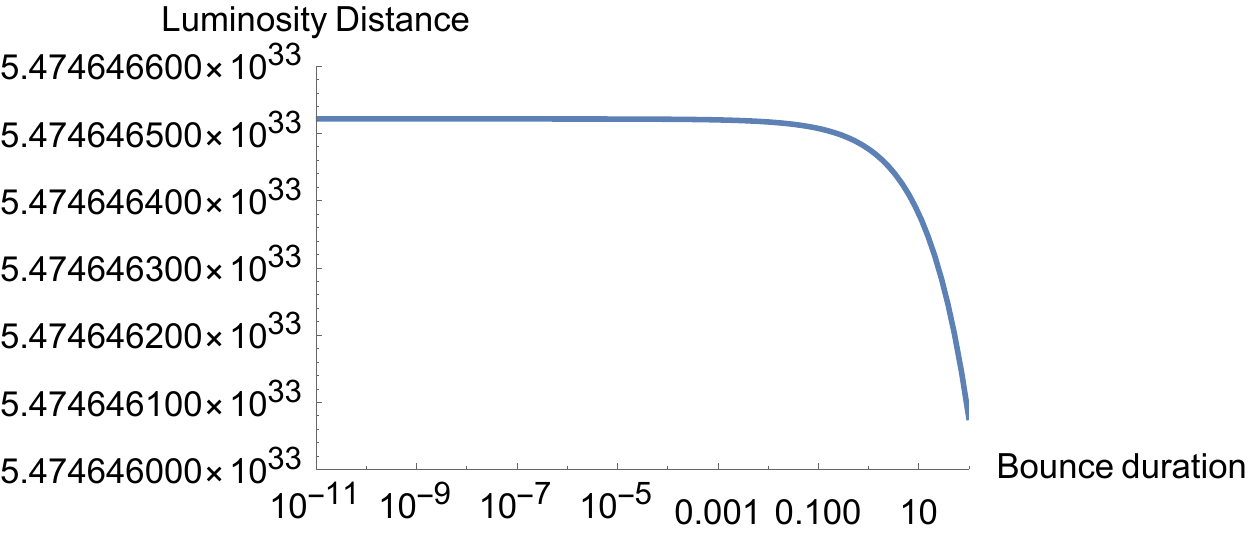}
\caption{Luminosity distance (m) as a function of the bounce duration (s) between an event in the contracting phase and the current universe (including radiation dominated and matter matter dominated phases).}
\label{bdur}
\end{figure}

In Fig. \ref{bdur} we consider the luminosity distance between an event in the contracting phase and the contemporary universe, as a function of the ``bounce duration". Fig. \ref{bdur} shows that the detailed value of $t_B$ does not care in the following analysis: the contribution of the bounce phase to the full integral is negligible. 

\begin{figure}[H]
\includegraphics[width=85mm,center]{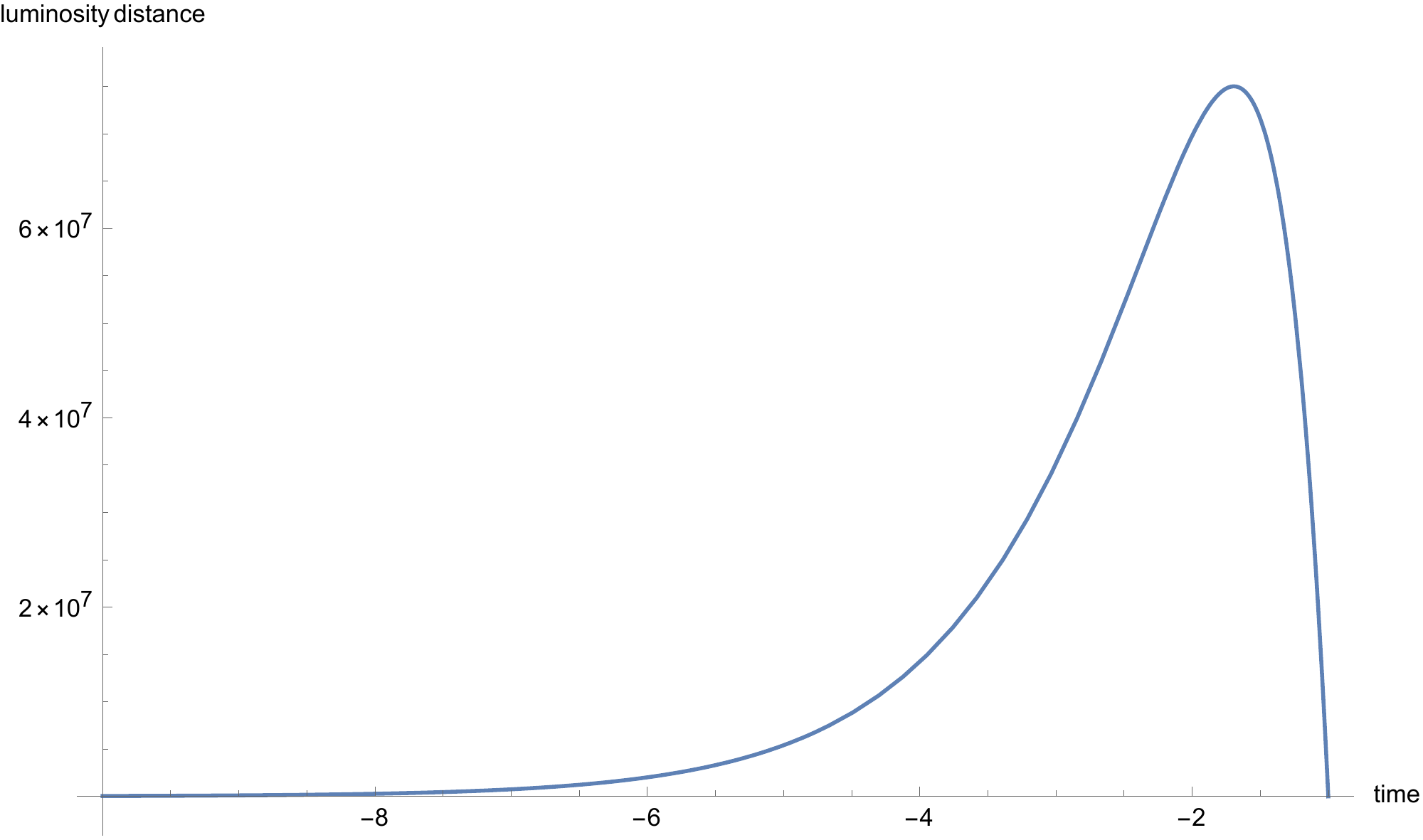}
\caption{Luminosity distance (m) as a function of the emission time $t_e$ (s) in the contracting branch for an exponential contraction. The reception time $t_r$ has been set to 1 second before the bounce and $\alpha$ was arbitrarily set to 1 in order to increase the readability.}
\label{expdl}
\end{figure}

Finally, it is worth considering the cosmological constant case, $a(t)=ke^{-\alpha t}$, where $\alpha=|H|>0$. The luminosity distance then reads
\begin{equation}
D_L=c\frac{e^{\alpha (t_e -2t_r)}}{\alpha}\left[ e^{\alpha t_r}-e^{\alpha t_e} \right].
\label{DL6}
\end{equation}
Clearly, in this case again, $D_L\rightarrow 0$ when $t_e\rightarrow -\infty$, as illustrated in Fig. \ref{expdl}. Sources located in the remote past have their flux intensely amplified.

\section{Consistency conditions}

The results given in the previous section do raise some questions. The case $n>1/2$ is in no way exotic from the point of view of the equation of state. It actually corresponds to a usual matter dominated universe, as naively expected far away from bounce. The behavior of the luminosity distance is then such that sources that have emitted light in an arbitrary distant past will lead to a measured flux which is arbitrarily amplified by the contraction of the scale factor. This basically means that the energy density will diverge at all points in space, leading to a kind of {\it new Olbers paradox} worsened by the contraction. In addition the frequency will also become arbitrarily high. As a consequence, the Universe cannot have been forever in a contraction phase with $n>1/2$ and filled with objects emitting energy. The energy density growth would anyway trigger the bounce -- at least in quantum-gravity models where the energy density is bounded from above by quantum geometry repulsive effects. This consistency condition has to be taken into account when building a consistent bouncing universe. 

The case $n=1/2$ is not fundamentally different. The luminosity distance being nearly constant, the energy amount received by each space point would also diverge in a forever-contracting universe. It should be pointed out that even for $n<1/2$ the space integral of any homogeneous source term will obviously diverge, as this is already the case in a static Minkowski universe.\\

The exponential contraction case is slightly more subtle. The luminosity distance is rapidly going to zero. The amplification due to the fast contraction of the Universe is thus very intense. However the  horizon and physical distances relative evolutions are such that the comoving Hubble radius is shrinking when going backward in time in the contracting branch (as when going forward in time in the expanding branch). The number density of sources causally linked to any space point will therefore also tend to zero and eventually solve the apparent paradox. 

\section{Seeing through the bounce}

Those considerations raise the important question of the possible observation of events having taken place before the bounce. Obviously, most signals or objects possibly existing in the contracting branch will be destroyed of washed out by the huge density reached -- in most models -- around the bounce time. The only exception could be gravitational waves. This is the only signal coupled weekly enough to matter so that it could propagate through the bounce (the details depend on the specific model considered). This has been investigated in different cases (see, {\it e.g.}, \cite{Gurzadyan:2013cna,Nelson:2011gb}) but focusing only on geometrical aspects -- ignoring the aforementioned amplification -- and considering consequences on the cosmological microwave background (CMB) spectra.\\

Let us consider here a different scenario. The hypothesis is that an event emitting intense gravitational waves has taken place before the bounce, {\it e.g.} the coalescence of two massive black holes (BHs). Clearly we don't know what the Universe looked like before the bounce. We however assume here that events comparable to what happens in our expanding branch took place in the contracting branch. At the lowest order the wave amplitude produced by a binary system and observed far away can be written \cite{Maggiore:1900zz}:

\begin{equation}
h=\frac{4}{D_L}\left[ \frac{G\mathcal{M}}{c^2} \right]^{\frac{5}{3}}\left[ \frac{\pi f}{c} \right]^{\frac{2}{3}}g(\tau,\Phi(f)),
\label{h}
\end{equation}
where $\mathcal{M}$ is the chirp mass, $f$ is the gravitational wave frequency at the observer location, $g$ is a sum and product of trigonometric functions (different for different polarisations) depending on $\tau$, the angle of the orbital plane, and on the phase $\Phi(f)$.\\

\begin{figure}[H]
\includegraphics[width=85mm,center]{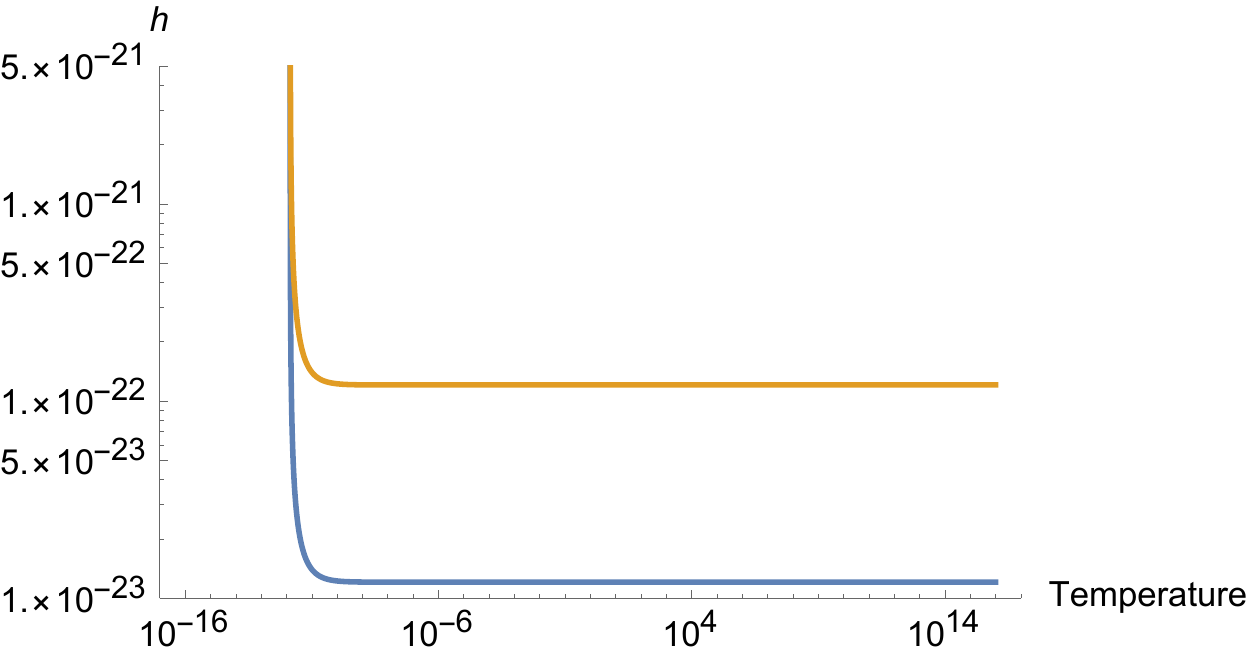}
\caption{Gravitational wave amplitude today as a function of the bounce temperature (GeV). The upper curve is for $10^9$ solar masses BHs and the lower curve for $10^8$ solar masses BHs, both merging in the contracting phase.}
\label{h1}
\end{figure}

As quite a lot of bouncing models are justified as alternatives to inflation (although bounces are compatible with inflation \cite{lqc2}), it is instructive to focus on a non-inflationary scenario and to study whether a pre {\it Big Bounce} signal can be detected. (An inflationary phase would obviously dilute the signal to a vanishingly small amplitude.) We consider the following toy-model: a contracting radiation-dominated phase, followed by a stationary bouncing phase, followed by the usual radiation-dominated and matter dominated stages. The number of efolds between the bounce and today is of course a relevant parameter that we express through the temperature of the Universe at the bouncing time. On Fig \ref{h1}, we have plotted the amplitude of gravitational waves emitted by the coalescence of 100 millions and one billion solar masses BHs as a function of the bouncing temperature. Interestingly, for a radiation dominated contracting phase, because the luminosity distance rapidly reaches an asymptotic value, it is not necessary to specify the merging time as long as it is far enough before the bounce. As it can be noticed from the curves, as soon as the temperature is chosen at a reasonable value, the amplitude is constant and becomes non-negligible and comparable to the sensitivity of current or next-generation experiments. The $h$ asymptotic behavior -- which might appear as quite strange at first sight -- is just due to the converging property of the integral of $1/a$ which enters the definition of the luminosity distance.

\begin{figure}[H]
\includegraphics[width=85mm,center]{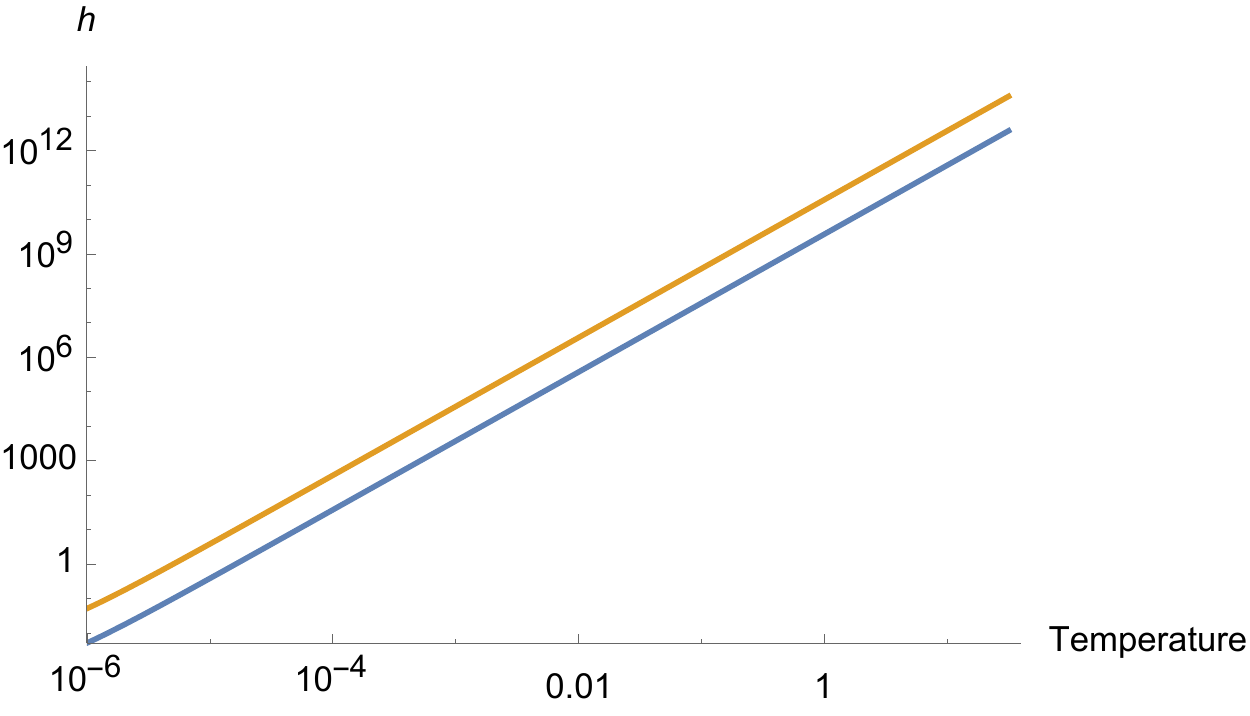}
\caption{Gravitational wave amplitude at the bounce time as a function of the bounce temperature (GeV). The upper curve is for $10^9$ solar masses BHs and the lower curve for $10^8$ solar masses BHs, both merging in the contracting phase.}
\label{h2}
\end{figure}

An obvious limitation of this calculation comes from the perturbative treatment. As it can be seen in Fig. \ref{h2}, as long as the bounce temperature is set much above the nucleosynthesis temperature, the gravitational waves amplitude at the bounce becomes too large to justify a perturbative calculation. This is a limitation to the presented study -- which requires a deeper treatment for this case -- but not to the presented idea in itself.\\

However, if the bounce temperature is set to the lowest possible one, the amplitude at the bounce is marginally compatible with a perturbative approach and this study shows -- in a consistent way -- that, in principle, gravitational waves from events occurring in the contracting phase of bouncing models could be detected in the contemporary universe.\\

One could also consider a phase on matter domination preceding the radiation dominated era in the contracting branch. If sources are located in this matter dominated phase, the amplitude does depend on the time at which the coalescence takes place. It is then possible to achieve nearly any value by choosing an emission time in the deep past. But the breakdown of the perturbative treatment would them become drastic and the whole result would be questionable. We therefore restrict ourselves to the radiation dominated case. \\

Another limitation is associated with the homogeneous and isotropic treatment of the bouncing universe. This should be considered as a toy-model  approximation. It is however not fully irrelevant. First, it should be pointed out that many bouncing models have been shown to resist the inclusion of anisotropies (see, {\it e.g.} \cite{Ashtekar:2009vc} for the case of loop gravity) with a quite minor modification of the Friedmann equation \cite{Linsefors:2013bua}. Anisotropic stress on gravitational waves could even be a way to discriminated between models. The homogeneous treatment is harder to justify and should obviously be seen as a first step. Recent calculations \cite{Clifton:2017hvg} have however shown that exact solutions describing a regular lattice of black holes in a cosmological bouncing background do exist.


\section{Conclusion}

In this article we have shown that the luminosity distance in a contracting universe has a highly non-trivial behavior. Because of the ``competition" between the expanding wave dilution and the amplification due to the decreasing scale factor, in some cases ($n>1/2$), the luminosity distance between two events in the contracting branch does {\it decrease} with an increasing time difference.\\

As a consequence, some violent events releasing gravitational waves and taking place in the contracting branch of the Universe could be detected today. The question of their frequency is hard to be answered unequivocally as it obviously depends on the precise emission time which, in the case $n=1/2$, has strictly no impact on the luminosity distance. We leave for a future study the associated statistical analysis, together with the systematic study of the characteristic signatures of ``pre-bounce" signals. 

It can already be underlined that several possible ways of discriminating between ``pre-bounce" events and usual ``post-bounce" events do exist. The most obvious approach is purely statistical: the number of events should simply be higher than expected if sources located before the bounce contribute to the measured events. Beyond this obvious statement, one should also look for the absence of electromagnetic counterparts. Although not demonstrated, electromagnetic signals are usually expected to be associated with merging supermassive BHs. Thirdly, the measured luminosity distances for some events should lie outside of the usual range (either too large or to small). Finally, the measured luminosity distance (inferred from the frequency, the frequency evolution and the amplitude, see {\it e.g.} \cite{Holz:2005df} or \cite{Schutz:1997bw}) might mismatch the real one in a way which is observationally measurable.

\section{Acknowledgments}

K.M is supported by a grant from the CFM foundation.

\bibliography{refs}
 \end{document}